\documentclass{article}

     \PassOptionsToPackage{numbers, compress}{natbib}

 \usepackage[preprint]{neurips_2019}

\usepackage[utf8]{inputenc} 
\usepackage[T1]{fontenc}    
\usepackage{hyperref}       
\usepackage{url}            
\usepackage{booktabs}       
\usepackage{amsfonts}       
\usepackage{nicefrac}       
\usepackage{microtype}      
\usepackage{graphicx}
\usepackage{multirow}
\usepackage{multirow}
\usepackage{siunitx}
\usepackage{subcaption}
\usepackage[toc,page]{appendix}
\usepackage{booktabs}
\newcommand{\bftab}{\fontseries{b}\selectfont}

\title{Approach to Learning Generalized Audio Representation Through Batch Embedding Covariance Regularization and Constant-Q Transforms}

\author{ %
  Ankit Shah \\
  Language Technologies Institute\\
  Carnegie Mellon University\\
  Pittsburgh, PA 15213 \\
  \texttt{aps1@andrew.cmu.edu} \\
  \And
  Shuyi Chen \\
  Heinz Information College\\
  Carnegie Mellon University\\
  Pittsburgh, PA 15213 \\
  \texttt{shuyic@andrew.cmu.edu} \\
    \And 
    Kejun Zhou \\
  Heinz Information College\\
  Carnegie Mellon University\\
  Pittsburgh, PA 15213 \\
  \texttt{kejunz@andrew.cmu.edu} \\
  \And
  Yue Chen \\
  Heinz Information College\\
  Carnegie Mellon University\\
  Pittsburgh, PA 15213 \\
  \texttt{yuechen2@andrew.cmu.edu} \\
  \And 
  Bhiksha Raj\\
  Language Technologies Institute \\
  Carnegie Mellon Institute \\
    Pittsburgh, PA, 15213 \\ 
    \texttt{bhiksha@cs.cmu.edu} \\ 
}

\begin{document}

\maketitle

\begin{abstract}
General-purpose embedding is highly desirable for few-shot even zero-shot learning in many application scenarios, including the audio tasks. In order to understand representations better, we conducted thorough error analysis and visualization of HEAR 2021 submission results. Inspired by the analysis, this work experiments with different front-end audio preprocessing methods, including Constant-Q Transform (CQT) and Short-time Fourier transform (STFT), and proposes a Batch Embedding Covariance Regularization (BECR) term to uncover a more holistic simulation of the frequency information received by the human auditory system. We tested the models on the suite of HEAR 2021 tasks, which encompass a broad category of tasks. Preliminary results show (1) the proposed BECR can incur a more dispersed embedding on the test set, (2) BECR improves the PaSST model without extra computation complexity, and (3) STFT preprocessing outperforms CQT in all tasks we tested. \\
\textbf{Github: } \href{https://github.com/ankitshah009/general_audio_embedding_hear_2021}{https://github.com/ankitshah009/general\_audio\_embedding\_hear\_2021} \\
\end{abstract}

\section{Introduction}
General-purpose representation learning is still an open question in audio datasets. Therefore, Holistic Evaluation of Audio Representations 2021 (HEAR 2021) challenge was proposed, aiming at providing longitudinal insights into different generalized audio representation models \cite{HEAR}. The challenge was to train one audio representation model that is flexible enough to represent unseen audio datasets. The representations were evaluated by training and testing a shallow network built on the embedding output of the models. The end-to-end process of HEAR2021 is summarized in Fig.~\ref{figure1}.

Inspired by the results of the challenge, this work first compares Short-time Fourier transform (STFT) with Constant-Q Transform (CQT), which was not used by any team in the challenge as the audio preprocessing method. Secondly, based on thorough error analysis and visualization of HEAR 2021 submission results, we propose Batch Embedding Covariance Regularization (BECR), a regularizing term that utilizes the Gini Index to measure the statistical dispersion of eigenvalues of the covariance matrix of the embedding on a training task. More specifically, it encourages the projection of representations of a specific pre-training task in all its eigenvectors to be as evenly dispersed as possible. Therefore, it aims to learn a deep representation network that is more versatile in low-dimension space when trained with only one dataset of a specific domain. We also propose an optimized implementation algorithm to reduce the time complexity.

We tested the two proposals along with a baseline model on four HEAR 2021 tasks, which encompass tasks from three audio domains including speech, music, and broad. Results show BECR improves the PaSST baseline in all tasks while CQT-trained results are inferior compared to Mel STFT models.

\section{Related Work}
\subsection{Audio Data Preprocessing Techniques}
\subsubsection{Short-time Fourier transform (STFT) and Mel Spectrogram}
An approach to better solve the general representation learning challenge is by applying different hand-crafted transformations based upon domain expertise for different tasks, for example, using Short-time Fourier transform (STFT) \cite{HEAR}. STFT is a powerful audio signal processing tool that can be applied in many tasks. It specifies complex amplitude versus time and frequency for every signal and defines a valuable class of time-frequency distributions. However, the STFT has its disadvantages, such as the limit in its time-frequency resolution capability. Low frequencies can be hardly depicted with short windows, whereas short pulses can only poorly be localized in time with long windows.\cite{ucdavis} As humans don’t perceive the sound in linear scale, Mel scale is proposed such that equal distances in pitch are equally distant to the listener, the human. The Mel spectrogram converts the values in hertz to Mel scale. This transformation can better stimulate human hearing than STFT \cite{ZShi}.  

\subsubsection{Constant Q-transform (CQT)}
Another way suitable to preprocess the music, human voice, and other sound varying data is Constant Q-transform. In 1991, Brown proposed CQT to simulate the human auditory system by using a transform with a fixed quality factor Q \cite{DWang}. Constant Q-transform is different from STFT in several ways. The Constant-Q transform has logarithmically spaced frequency bins, while the frequency component of STFT is linear. Further, the Constant-Q transform has octaves bin widths other than absolute value bin width. As the output of the transform is effectively amplitude/phase against log frequency, fewer frequency bins are required to cover a given range effectively \cite{ZShi}. Therefore, some argue Constant-Q transform better describes what is received by the human auditory system and is thus better positioned in the musical area \cite{Christian}. 

\subsection{Gini Index in Machine Learning}
Gini Index is a data purity measure. A small Gini Index indicates a high purity of the encodings or signals. Gini Index has been widely applied in Machine Learning. For example, Randall \cite{Balestriero} proposed neural decision trees (NDT) based on decision trees and MLP in the practice of combining Gini Index with neural networks. Park \cite{Park} also proposed a deep learning model using the Gini Index algorithm for the extraction of features from datasets. 

We noticed the need for a summary statistic that describes the overall geometric property of the embedding matrix on an evaluation dataset during the analysis of HEAR2021 results -- specifically, how spread out the embedding for different tasks. Therefore we got the idea to apply Gini Index to normalized eigenvalues of embedding matrix as a regularization term. We believe this work presents the first application of Gini Index with such a definition in audio tasks.


\section{Method}
\label{Section3}

\begin{figure}[ht]
\centering
\includegraphics[width=0.7\textwidth]{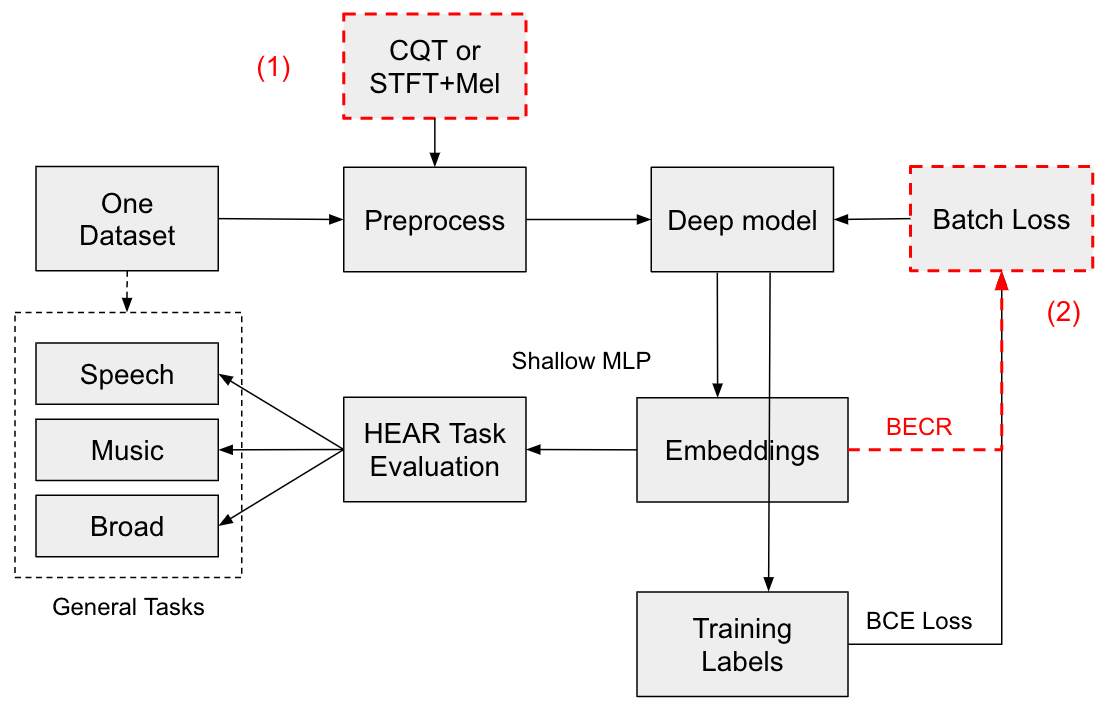}

\caption{The End-to-end Training and Evaluation Process for HEAR2021. In this work, we made two modifications of it. (1) We compared the effects of CQT and STFT, and (2) we designed a regularization term in the training time.}
\label{figure1}
\end{figure}

The end-to-end process of this work is summarized in Fig.~\ref{figure1}. We first compare the effects of two preprocessing methods. Secondly, we experiment with a novel Gini Index-based regularization to improve the versatility of the model. The resulting models are used to generate embedding on a variety of unknown datasets of HEAR2021 dataset, which will be used to train shallow MLP layers to get a final evaluation score. 

\subsection{Preprocessing Methods}

Short-time Fourier transform (STFT) first divides the long recording signal into short equal segments in the time domain. Then, it computes a Fourier transform on each segment and generate several frequency spectrums. Discrete STFT can be expressed as:

\begin{equation}
    X(m,\omega) = \sum^{\infty}_{n=-\infty}{x[n]\omega[n-m]e^{-j\omega n}}
\end{equation}

CQT transform mirrors the human auditory system, whereby at lower-frequencies spectral resolution is better \cite{wikiCQT}. Discrete CQT can be expressed as: 

\begin{equation}
    X[k] = \frac{1}{N[k]}\sum^{N[k]-1}_{n=0}{W[k,n]x[n]e^{\frac{-j2\pi Q n}{N[k]}}}  
\end{equation}

\subsection{Baseline Architecture}
\begin{figure}[h]
\centering
\includegraphics[width=0.5\textwidth]{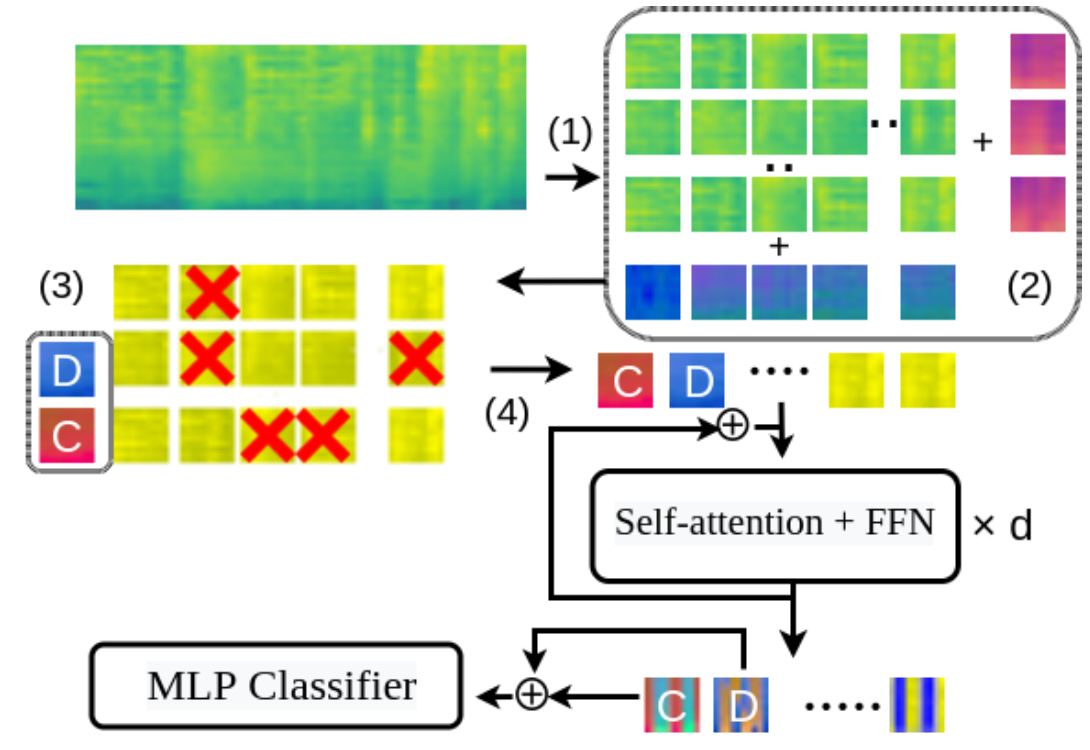}
\caption{The Patchout transformer (PaSST) architecture \cite{Kou}.}
\label{figure2}
\end{figure}

We choose PaSST model as the baseline structure \cite{Kou} which achieved overall top results in the 19 tasks (14 are secret tasks) of HEAR2021 challenge \cite{HEAR}. PaSST is the state-of-the-art transformer-based audio model that can achieve SOTA results with less memory and time complexity compared to other CNN-based models \cite{Kou}. As shown in Fig.~\ref{figure2}, the input of the model is an audio spectrogram generated by preprocessing methods. In part 1, it experiences a patch extraction and linear projection. In part 2, frequency and time positional encodings are added. Then a Patchout operation will be applied. The Patchout idea is to encourage the transformer to classify with an incomplete sequence, similar to dropout. Finally, the sequence is flattened and then passed through Self-attention layer. In the last, a classifier MLP layer operates on the classiﬁcation token and generates predictions \cite{Kou}. 

\subsection{Batch Embedding Covariance Regularization (BECR)}
\subsubsection{Analysis of Low-dimension Representation Projections}
\label{sec:3.3.1}
\begin{figure}
     \centering
     \begin{subfigure}[b]{0.46\textwidth}
         \centering
         \includegraphics[width=\textwidth]{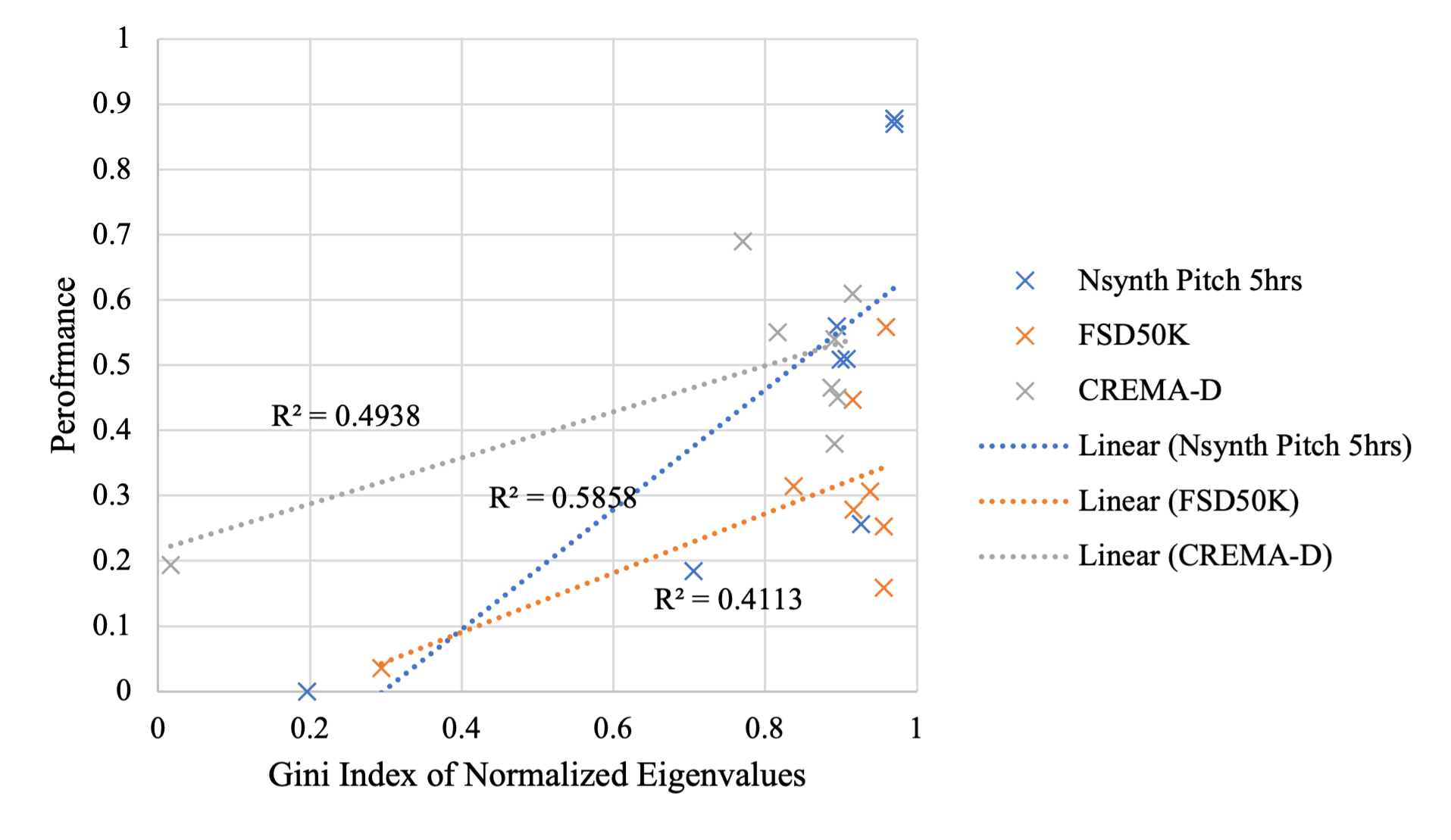}
         \caption{Gini Index of normalized eigenvalues}
     \end{subfigure}
     \hfill
     \begin{subfigure}[b]{0.46\textwidth}
         \centering
         \includegraphics[width=\textwidth]{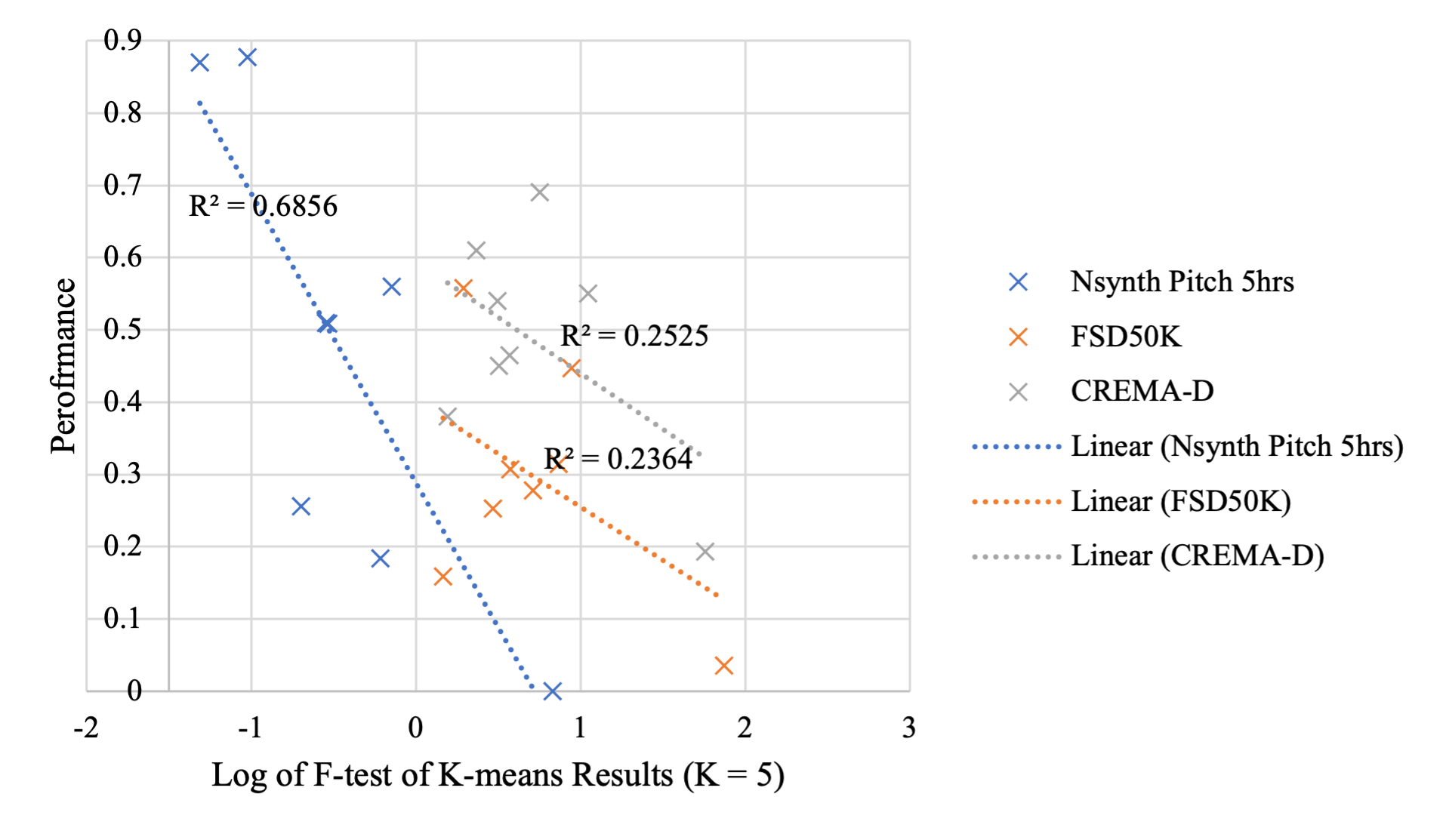}
         \caption{F-test of K-means results}
     \end{subfigure}
     \vskip\baselineskip
     \begin{subfigure}[b]{0.46\textwidth}
         \centering
         \includegraphics[width=\textwidth]{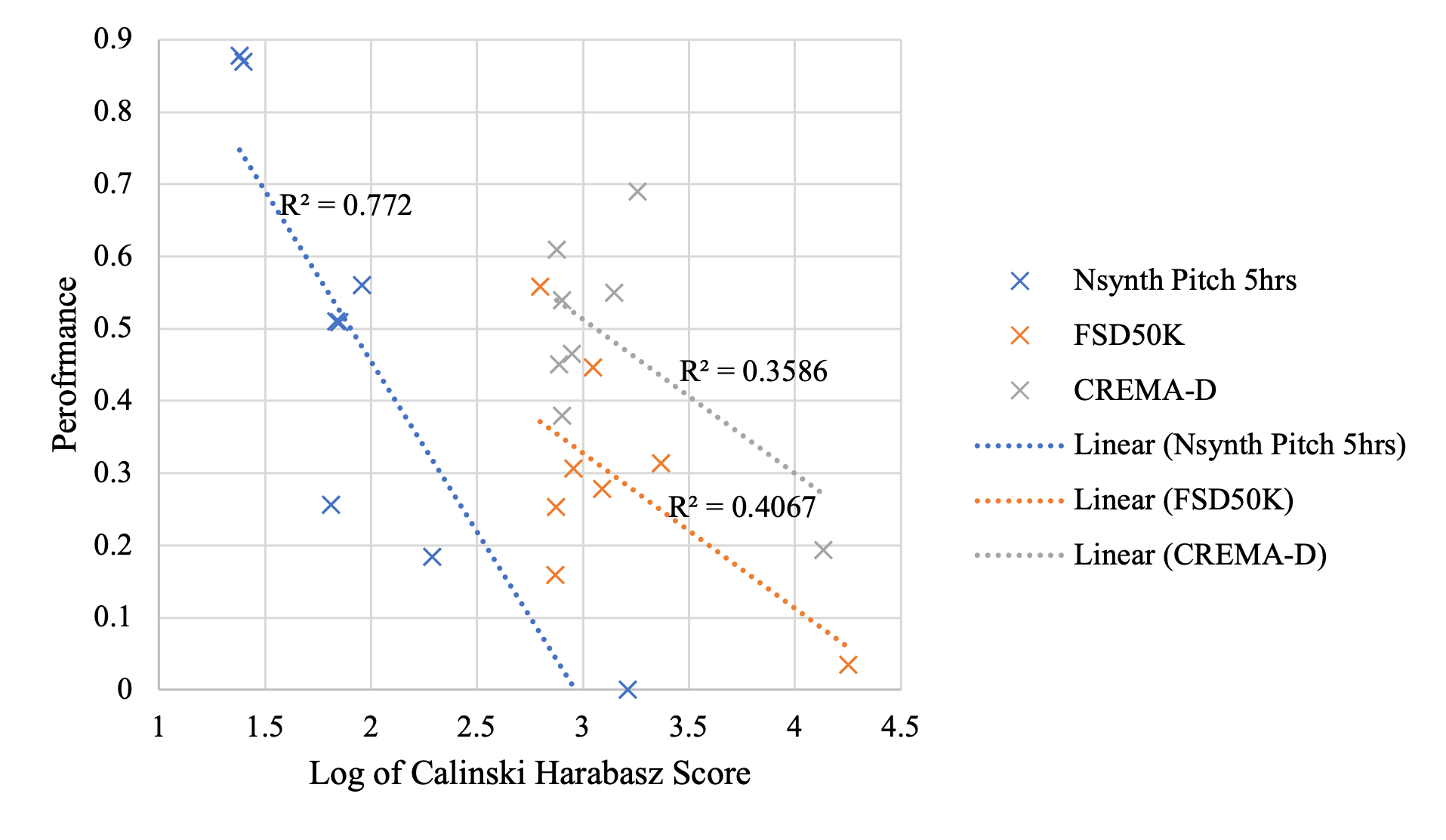}
         \caption{Calinski-harabasz score of K-means results}
     \end{subfigure}
     \hfill
     \begin{subfigure}[b]{0.46\textwidth}
         \centering
         \includegraphics[width=\textwidth]{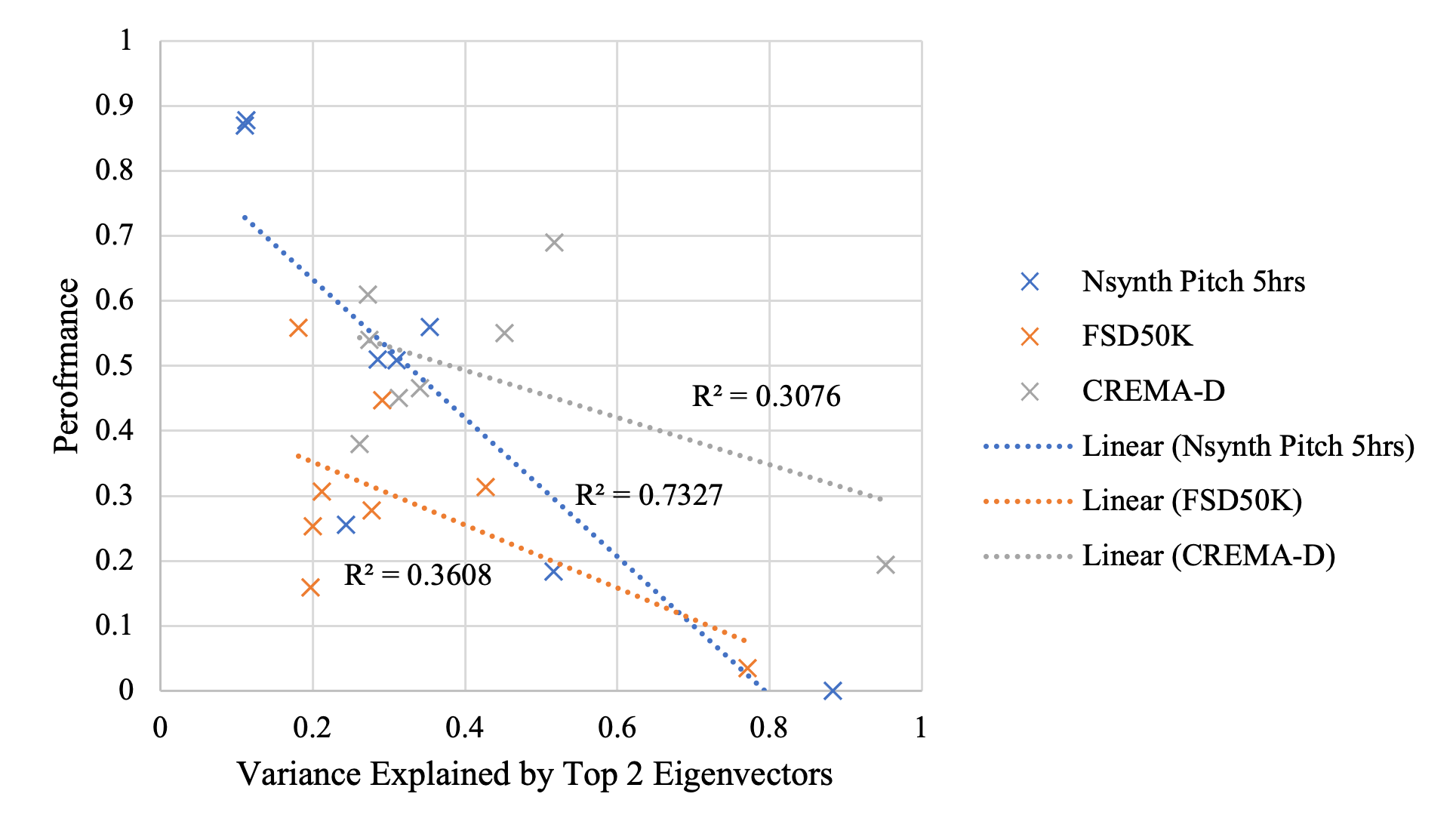}
         \caption{Variance explained by top 2 principal vectors}
     \end{subfigure}
        \caption{Summary of Embedding Performance and Dispersion Metrics. Each data point is either a submitted or a replicated model of HEAR2021. Nsynth Pitch, FSD50K, and CREMA-D are music, broad, and speech tasks respectively.}
        \label{figure3}
\end{figure}

Through analysis of the embedding of top models in various tasks, we observed that given the same task, those models that perform better in the downstream task generally have higher embedding dispersion, for example with lower variance explained by top principal components and lower K-means F-test score. See Fig.~\ref{figure3} for a summary. Thus, we conjecture the high dispersion of the embedding produced by a model is helpful in downstream tasks. 

\subsubsection{BECR Design}
Since we have restricted the training process to only using one task dataset, the eigenvalues of the covariance matrix are suitable to simulate the variance of embedding when projected to different dimensions for different testing tasks. The Gini Index of normalized eigenvalues is therefore a summary statistic that describes how evenly dispersed the embedding is across the eigenspace. So we define BECR in the following way:

For embedding layer with $D$ outputs, $\mathbf{K}$ is the $D$ dimensional covariance matrix of each batch embedding $f_\theta(\mathbf{X}_i)$,

\begin{equation}
\mathbf{K}(f_\theta(\mathbf{X}_i))=(f_\theta(\mathbf{X}_i)-\mathrm{E}[f_\theta(\mathbf{X}_i)])(f_\theta(\mathbf{X}_i)-\mathrm{E}[f_\theta(\mathbf{X}_i)])^{\mathrm{T}}
\end{equation}

$\mathbf{K}$ is always positive semi-definite with real nonnegative eigenvalues. $\mathcal{G}$ applies the definition of Gini Index to the normalized eigenvalues of $\mathbf{K}$, 

\begin{equation}
\mathcal{G}(f_\theta(\mathbf{X}_i)) = 1-  \sum_{i = 1}^{D} (\frac{\lambda_i(\mathbf{K}(f_\theta(\mathbf{X}_i)))}{\sum_{i=1}^{D} \lambda_i(\mathbf{K}(f_\theta(\mathbf{X}_i)))})^2
\label{eq:4}
\end{equation}

$\mathcal{R}$ is the proposed regularization term,

\begin{equation}
\mathcal{R}(\mathbf{X}_i, \theta) =  \max(0, \epsilon - \mathcal{G}(f_\theta(\mathbf{X}_i)))^2
\label{eq:5}
\end{equation}

where $\epsilon$ is a hyperparameter, defining the upper bound of the Gini Index when incurring a loss.

Finally, the total loss is defined by
 
\begin{equation}
\mathcal{L'}(\mathbf{X}_i, \theta) = (1-\lambda) \mathcal{L}(\mathbf{X}_i, \theta) + \lambda \mathcal{R}(\mathbf{X}_i, \theta)
\end{equation}

In our case, the vanilla loss $\mathcal{L}$ is Binary Cross Entropy loss, whose range is [0, 1]. By adding the regularization term, it encourages a large Gini Index that is encouraging evenly distributed eigenvalues. Also, the regularization term is a convex function added to the original loss function, which has desirable convergence property.

\subsubsection{Implementation Details of BECR}

Let batch size be N and embedding dimension be K. The eigenvalue decomposition algorithm takes $O(K^3)$ complexity. Adding the covariance matrix calculation, the total additional complexity per batch is $O(K^3 + K^2 * N)$, which is categorically impractical in our case since N is around 10 and K is around 1000. Through our experiment, training one 10-sample batch of FSD50K dataset with embedding eigenvalue decomposition takes around 6 hours in an RTX 3090 machine, 18 times longer than that without the loss. Therefore, we propose an optimized implementation without eigenvalue decomposition, that is with a complexity of $O(K^2 * N)$.

Recall $tr(A) = \sum_i \lambda_i(A)$ and $tr(A^2) = \sum_i \lambda_i(A)^2$. Therefore, Eq.~\ref{eq:4} can be simplified to

\begin{equation}
\mathcal{G}(f_\theta(\mathbf{X}_i)) = 1-  \sum (\frac{\lambda_i}{\sum \lambda_i})^2
=1- \frac{\sum \lambda_i^2}{(\sum \lambda_i)^2} = 1-\frac{tr(\mathbf{K}(f_\theta(\mathbf{X}_i))^2)}{tr(\mathbf{K}(f_\theta(\mathbf{X}_i)))^2}
\end{equation}

Through experiment, the speed of this implementation method is similar to vanilla loss calculation. Simplified BECR takes 33 hours compared to vanilla loss's 36 hours in training as shown in Table~\ref{result}. So we can apply BECR with little extra complexity.

\section{Experimental Evaluation}
\label{gen_inst}
\subsection{Datasets and Evaluation Metric}\

We evaluate the performance of the models on three types of data sets: music, speech, and broad sounds. See Table~\ref{dataset} for dataset and evaluation metrics summary.

For music, we use the NSynth Pitch containing 305,979 musical notes, each with a unique pitch, timbre, and envelope. For 1,006 instruments from commercial sample libraries, the dataset was generated in four seconds, monophonic 16kHz audio snippets by ranging over every pitch of a standard MIDI piano (21-108) as well as five different velocities (25, 50, 75, 100, 127). The goal of this task is to classify instrumental sounds into one of 88 pitches \cite{NSynth}. The Pitch Accuracy was used for evaluation on NSynth Pitch task.

We also use Beijing Opera Percussion for music task evaluation. Beijing Opera Percussion Instrument Dataset is based on recordings from The Beijing Opera \cite{beijingopera}. It contains six main percussion instruments that can be classified into four main categories: Bangu, Naobo, Daluo, and Xiaoluo. There are 236 audio clips in total. Classification accuracy is used for evaluation on Beijing Opera Percussion task.

For speech, we use the CREMA-D for emotion recognition \cite{cao}. This dataset contains audio data of actors reciting sentences with one of six different emotions (anger, disgust, fear, happy, neutral, and sad). The goal of this task is to identify the type of emotion the actors are in when they say the sentences. Classification accuracy is used for evaluation on CREMA-D task.

For broad sounds, we use the FSD50K, each of the audio clips in this dataset is labeled using one or more of 200 classes in environmental sounds, speech, and music \cite{fonseca}. This dataset contains over 51K audio clips, totaling over 100 hours of audio, and is extracted from the AudioSet Ontology. We also use FSD50K for training. mAP was used for multi-label evaluation on FSD50K task.

\begin{table}[ht]
\centering
  \begin{tabular}{p{2.5cm}p{2.5cm}p{2.5cm}p{2.5cm}p{4.5cm} } 
    \toprule
    Task Name & Predictor Type & Split Mode & Evaluation Metric \\
     \midrule
     NSynth Pitch 5h  & C & TVT & Pitch Acc. \\
     Beijing Opera  & C & 5-fold & Accuracy \\
     CREMA-D & C & 5-fold & Accuracy \\
     FSD50K & L & TVT & mAP \\
    \bottomrule
    \hfill
  \end{tabular}
  \caption{Summary of the four evaluation tasks selected from HEAR 2021 \cite{HEAR}. For all four tasks, the embedding type are either scene based. The predictor types are either multiclass (C) or multilabel (L). The split method used during downstream evaluation are either train/validation/test (TVT) or K-fold.}
  \label{dataset}
\end{table}

\subsection{Hyperparameter Tuning}
\label{sec:1}

The proposed BECR involves two hyperparameters, $\lambda$ and $\epsilon$. We only experimented with $\lambda$ of 0.05 and 0.10 as through our experiment $\lambda \le 0.1$ ensures the BECR does not cannibalize all the loss in the beginning few epochs. 

For determination of search space of $\epsilon$, we observed that the Gini Index of vanilla PaSST's embedding on FSD50K is 0.92 after 100 epochs (See Table~\ref{comparison}). Also, in the initial batches, the Gini Index is around 0.3-0.6 in experiments. So it would not make sense if we set epsilon smaller than 0.6 which makes little difference in the final output (See Eq.~\ref{eq:5}). So we experimented with $\epsilon$ larger than 0.7. The tuning results in Table~\ref{tuning} show $(\lambda,
\epsilon) = (0.05, 0.7)$ is the best combination.

\begin{table}[ht]
\centering
  \begin{tabular}{p{2cm}p{2cm}p{2cm}p{2cm}p{4cm} } 
    \toprule
$\epsilon$ & $\lambda$ & \# Epochs & Test Set Performance \\
     \midrule
     0.7 & 0.05& 50 (19hr) & \textbf{30.7} \\
     0.8 & 0.05& 50 (18hr) & 24.4\\
     0.9 & 0.05& 50 (18hr) & 24.9\\
     0.7 & 0.10& 50 (19hr) & 25.6\\
     0.8 & 0.10& 50 (18hr) & 25.7\\
     0.9 & 0.10& 50 (18hr) & 29.8\\
    \bottomrule
  \hfill
  \end{tabular}
  \caption{Hyperparameter Tuning Results on Training Set (FSD50K)}. 
  \label{tuning}
\end{table}

\section{Results}
Results show that CQT-preprocessing is worse a choice than STFT+Mel in all four tasks. Additionally, the computational complexity of CQT transformation is larger than STFT+Mel, taking more than two times than original STFT+Mel (approximately 1 hour per epoch in FSD50K with a batch size of 6). The proposed BECR combined with Mel+STFT outperforms the baseline model in all four tasks with similar training complexity. 

\begin{table}[ht]
\centering
  \begin{tabular}{lllllll}
    \toprule
    \multirow{2}{*}{\textbf{Models}} & \multirow{2}{*}{\textbf{\# Epochs}} &
      \multicolumn{4}{c}{\textbf{Downstream Evaluation Scores}  (\%)} \\
      &{}& {Beijing Opera} & {Nsynth Pitch 5} & {CREMA-D} & {FSD50K} \\
      \midrule
     STFT Mel+PaSST (Baseline) & 100 (33hr) & 90.6 & 50.9 & 46.5 & 27.8 \\
     STFT Mel+PaSST (BECR) & 100 (36hr) & \textbf{92.7} & \textbf{51.2} & \bftab{47.6} & \bftab{36.8} \\
     CQT+PaSST & 50 (50hr) & 36.8 & 4.8 & 19.4 & 3.5 \\
     \midrule
     CP-JKU PaSST in HEAR2021 \cite{HEAR} & Unknown & 96.6 & 25.6 & 61.0 & 55.8 \\
    \bottomrule
  \hfill
  \end{tabular}
  \caption{Summary of Model Performances. $\epsilon$ and $\lambda$ of BECR are tuned on training set in section \ref{sec:1}.}
  \label{result}
\end{table}



\subsection{Discussion on CQT's Results}
We tried to find some explanation for CQT's worse performance than STFT.

First, by comparing the resulting embedding of STFT and CQT-based PaSST model dimension reduction methods of PCA and T-SNE, we found projection of CQT are usually in linearly unseperatable shape, while those of STFT seems more cluster-like. Considering evaluation process builds a 2-layer MLP on the embedding results (see Fig~\ref{figure1}), it's reasonable to assume that embedding of CQT-based PaSST being unseperatable accounts for its bad performance in the downstream tasks.

\begin{figure}
     \centering
     \begin{subfigure}[b]{0.32\textwidth}
         \centering
         \includegraphics[width=\textwidth]{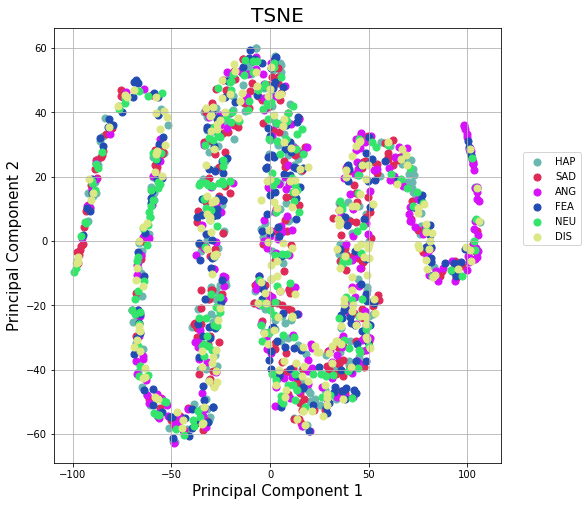}
         \caption{T-SNE of CQT-based PaSST Model Embedding}
     \end{subfigure}
     \hfill
     \begin{subfigure}[b]{0.32\textwidth}
         \centering
         \includegraphics[width=\textwidth]{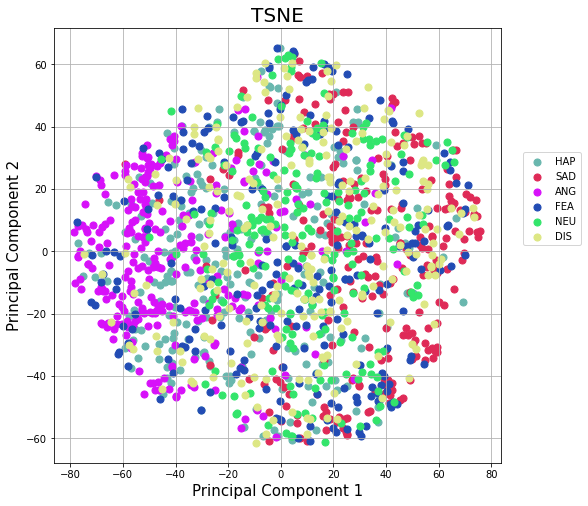}
         \caption{T-SNE of STFT-based PaSST Model Embedding}
     \end{subfigure}
     \hfill
     \begin{subfigure}[b]{0.32\textwidth}
         \centering
         \includegraphics[width=\textwidth]{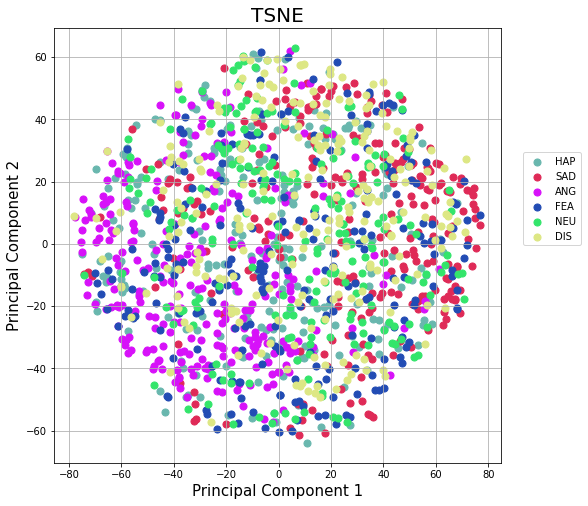}
         \caption{T-SNE of STFT-based PaSST with BECR Model Embedding}
     \end{subfigure}
     
      \begin{subfigure}[b]{0.32\textwidth}
         \centering
         \includegraphics[width=\textwidth]{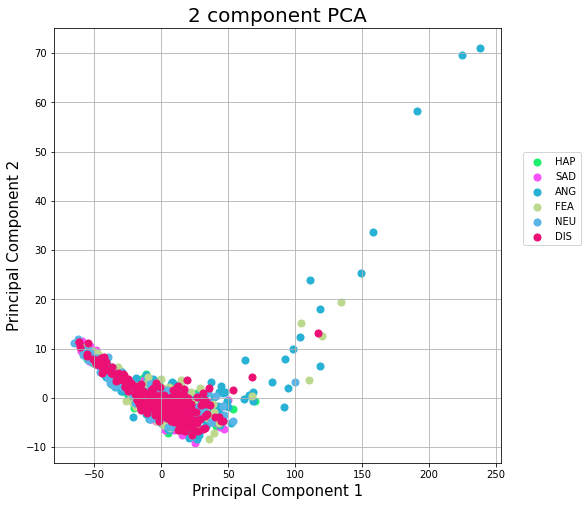}
         \caption{PCA of CQT-based PaSST Model Embedding}
     \end{subfigure}
     \hfill
     \begin{subfigure}[b]{0.32\textwidth}
         \centering
         \includegraphics[width=\textwidth]{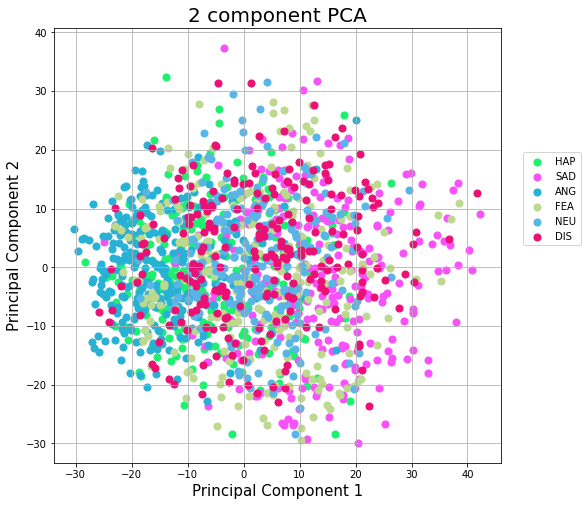}
         \caption{PCA of STFT-based PaSST Model Embedding}
     \end{subfigure}
     \hfill
     \begin{subfigure}[b]{0.32\textwidth}
         \centering
         \includegraphics[width=\textwidth]{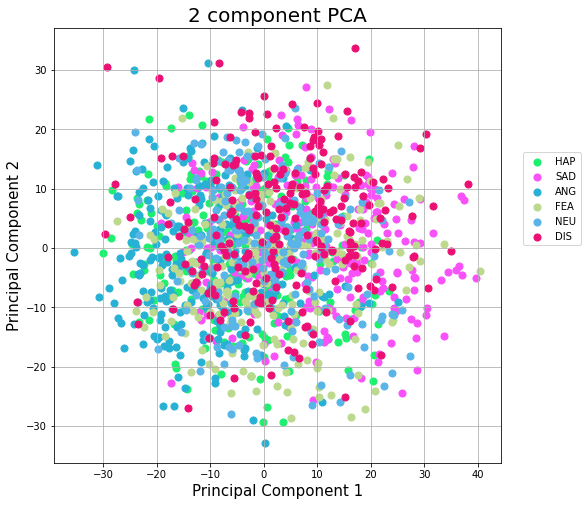}
         \caption{PCA of STFT-based PaSST with BECR Model Embedding}
     \end{subfigure}
        \caption{Projection of Embedding with PCA and T-SNE. The points are colored using the groundtruth labels. Dataset is CREMA-D. The models were trained on FSD-50K, not the CREMA-D dataset.}
        \label{figure4}
\end{figure}

Secondly, CQT-based PaSST performance and some of its metrics follow the pattern mentioned in Section \ref{sec:3.3.1}. Compared with STFT-based PaSST models, CQT-based PaSST embedding results are less dispersed in unseen datasets. See Table~\ref{comparison}.

Thirdly, the experimental results aside, we suspect a relatively "good" preprocessing method partly depends on the choice of model. The original implementation of PaSST uses STFT transformation \cite{Kou}, so it's natural that PaSST model works best with STFT preprocessing method instead of others.

\subsection{Discussion on BECR's Results}

\begin{figure}
     \centering
     \begin{subfigure}[b]{0.46\textwidth}
         \centering
         \includegraphics[width=\textwidth]{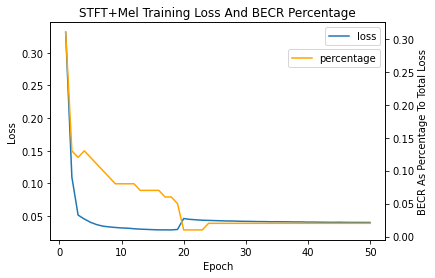}
         \caption{Total loss descent and BECR's percentage of total loss}
     \end{subfigure}
     \hfill
     \begin{subfigure}[b]{0.46\textwidth}
         \centering
         \includegraphics[width=\textwidth]{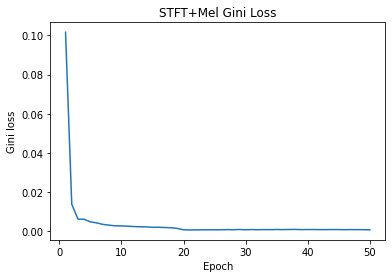}
         \caption{BECR term descent}
     \end{subfigure}
        \caption{BECR descent and the total training loss descent comparison ($\lambda = 0.05$, $\epsilon = 0.7$) }
        \label{figure5}
\end{figure}

\begin{figure}
     \centering
     \begin{subfigure}[b]{0.46\textwidth}
         \centering
         \includegraphics[width=\textwidth]{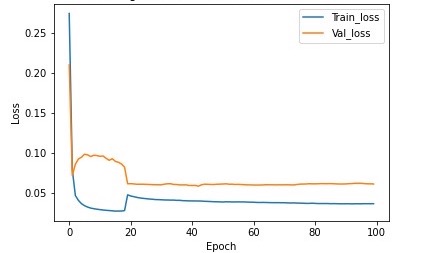}
         \caption{PaSST with STFT + Mel (Baseline)}
     \end{subfigure}
     \hfill
     \begin{subfigure}[b]{0.46\textwidth}
         \centering
         \includegraphics[width=\textwidth]{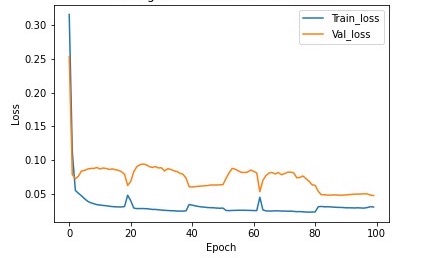}
         \caption{PaSST with BECR and STFT + Mel}
     \end{subfigure}
     \vskip\baselineskip
     \begin{subfigure}[b]{0.46\textwidth}
         \centering
         \includegraphics[width=\textwidth]{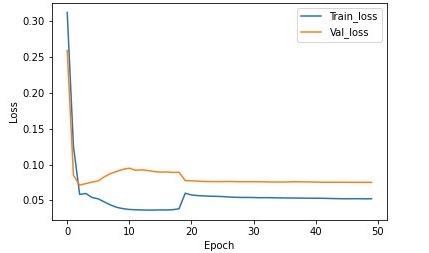}
         \caption{PaSST with CQT}
     \end{subfigure}
        \caption{Training loss descent of the models}
        \label{figure6}
\end{figure}

To verify that the BECR is making the improvement on baseline PaSST model, we show the regularization term successfully converges to zero after the first few epochs. See Fig~\ref{figure5}. In particular, the percentage of BECR term of total loss steadily descent from 15\% to 0\%, and does not cannibalize the total loss even in the early stage. Therefore compared with the baseline model, the validation loss descent is not significantly affected in the 100-epoch training process (Fig~\ref{figure6}).

Another piece of evidence is after adding BECR to the baseline PaSST, the variance explained by top eigenvectors decreases, F-test score decreases, and Gini Index of normalized eigenvalues increases, as BECR is intended for (see Table~\ref{comparison}). These changes indicate the embedding projection is more spread out than before, which leads to its better performance according to our analysis in Section \ref{sec:3.3.1}.

\begin{center}
\begin{table}
\renewcommand\arraystretch{1.3}
  \begin{tabular}{p{2.5cm}p{3.4cm}p{2.1cm}p{2.1cm}p{2.1cm} } 
    \toprule
    \textbf{Dataset} & \textbf{Models} & \textbf{Gini Index} & \textbf{Top 2 eigenvalues ratio} & \textbf{Performance}\\
 \hline
 \multirow{3}*{Nsynth Pitcht 5h}
 & CQT+PaSST    & 0.20   & 0.88 & 0.048\\
 & STFT+PaSST    & 0.90   & 0.31 & 0.509\\
 & STFT+PaSST (BECR)    & \textbf{0.91}    & \textbf{0.29} & \textbf{0.512}\\
                    
\multirow{3}*{FSD50K}
& CQT+PaSST    & 0.29   & 0.77 & 0.035\\
& STFT+PaSST    & 0.92    & 0.28 & 0.278\\
& STFT+PaSST (BECR)    & \textbf{0.94}    &\textbf{ 0.21} & \textbf{0.368} \\
                    
\multirow{3}*{CREMA-D}
& CQT+PaSST    & 0.02    & 0.95 & 0.195\\
& STFT+PaSST    & 0.89    & 0.34 & 0.465\\
& STFT+PaSST (BECR)    & \textbf{0.90}    & \textbf{0.31} & \textbf{0.476}\\
                    
\bottomrule
\end{tabular}
\caption{Effect of BECR. Models are trained on FSD50K dataset with 100 epochs of batch size 10.}
\label{comparison}
\end{table}
\end{center}

\section{Conclusion}

In this paper, we verify that CQT is a less ideal audio preprocessing method than STFT+Mel when used with PaSST model, not only decreasing downstream task performances but also dramatically increasing the computational complexity when transforming the audio data into the frequency domain. 

We also propose Batch Embedding Covariance Regularization (BECR), a Gini Index-based regularization term, which encourages widespread embedding in its eigenspace, and a fast implementation algorithm for it. We tested BECR with SOTA audio model PaSST in a wide variety of audio domains: music, speech, and broad and achieve better performance when compared with using PaSST alone trained with the same dataset and similar hours. The simplicity in intuition and low complexity of implementation to apply the method, together with the encouraging results in challenging unknown test tasks, demonstrate the promising potential BECR has for general-purpose audio representation learning.

We would like to note some limitations of this work. First, different models work well with different preprocessing methods. So the conclusion of CQT is better than STFT limits to our experiment setting which uses a PaSST model. Second, we have not verified if BECR is a generalizable technique working beyond PaSST model. It would be interesting to apply BECR to other common baselines in the future, for example, OpenL3 and wav2vec2 models, and see the difference it makes.

\section*{Acknowledgement}
Thanks to Chaoran Zhang, Yuxiang Zhang for their helpful comments on the work.

\bibliographystyle{IEEEbib}
\bibliography{IEEE}

\end{document}